\documentclass{emulateapj}
\submitted{Submitted to ApJ} 

\defcitealias{kmp+08}{K08}
\newcommand{\kms}{\,km\,s$^{-1}$}
\newcommand{\ergs}{\,ergs\,s$^{-1}$}
\newcommand{\msc}{G319.9--0.7}
\newcommand{\psr}{PSR~J1509--5850}

\begin{document}
\title{Radio Polarization Observations of \msc: A Bow-shock Nebula
with an Azimuthal Magnetic Field Powered by Pulsar J1509--5850}

\shorttitle{Radio Observations of \msc}
\shortauthors{Ng et al.}

\author{C.-Y. Ng\altaffilmark{1}, B. M. Gaensler\altaffilmark{2}, S. Chatterjee\altaffilmark{3}}
\affil{Sydney Institute for Astronomy, School of Physics, The University of Sydney, NSW 2006, Australia}
\author{\vspace*{-5mm}and\\S. Johnston}
\affil{Australia Telescope National Facility, CSIRO, P.O. Box 76, Epping, NSW 1710, Australia}
\email{ncy@physics.usyd.edu.au}

\altaffiltext{1}{Current address: Department of Physics, McGill
University, Montreal, QC, Canada H3A 2T8}
\altaffiltext{2}{ARC Federation Fellow}
\altaffiltext{3}{Current address: Department of Astronomy, Cornell University, Ithaca, NY 14853}

\begin{abstract}
We report radio polarization observations of \msc\ (MSC~319.9--0.7)
at 3 and 6\,cm obtained with the Australia Telescope Compact Array.
The source shows a highly elongated morphology with the energetic pulsar
J1509--5850 located at the tip. We found a flat radio spectrum of index
$\alpha=-0.26\pm0.04$ and a high degree of linear polarization. These 
results confirm \msc\ as a bow-shock pulsar wind nebula. The polarization
maps suggest a helical magnetic field trailing the pulsar, with the symmetry
axis parallel to the system's inferred direction of motion. This is the
first time such a field geometry has been seen in a bow-shock nebula, and
it may be the result of an alignment between the pulsar spin axis and its
space velocity. Compared to other bow-shock examples, \msc\ exhibits very
different properties in the field structure and surface brightness
distribution, illustrating the large diversity of the population.

\end{abstract}


\section{INTRODUCTION}
Pulsars lose a significant fraction of their rotational energy through
their relativistic winds. The consequent interactions with the ambient
medium result in broadband synchrotron emission. These structures are
collectively referred to as pulsar wind nebulae (PWNe). The properties of
a PWN depend strongly on its evolutionary state and environment. Since a
pulsar typically has a space velocity of a few hundred kilometers per
second, it will eventually escape its natal supernova remnant (SNR) and
travel supersonically in the interstellar medium (ISM). In such cases,
the pulsar outflow can be confined by ram pressure, resulting in a
bow-shock nebula. Multiwavelength observations have identified several
bow-shock PWN systems \citep[see review by][]{gs06}. The best-studied
example is `the Mouse' (G359.23--0.82), in which the X-ray and radio
emissions can be well-modeled by a bright head coincident with the
pulsar, a `tongue' region corresponding to the wind termination shock,
and an elongated tail associated with the post shock flow material
\citep{gvc+04}.

Studies of the Mouse and other bow-shock PWNe have set the stage for
subsequent theoretical and numerical modeling efforts. \citet{rcl05}
considered an axisymmetric case of a pulsar traveling along its spin
axis direction. They proposed an azimuthal $B$-field geometry for
bow-shocks, and derived the analytical shape of a pulsar tail. With
a similar aligned configuration, \citet{bad05} carried out the
first relativistic magnetohydrydynamic (MHD) simulations of bow-shocks to
study the effects of different wind magnetization. \citet{vmc+07} relaxed
the assumption of alignment and presented three-dimensional non-relativistic
hydrodynamic simulations to illustrate the dependence of bow-shock morphology
on pulsar orientation and ISM gradient.

Observationally, PWNe at radio frequencies are characterized by flat
spectra with spectral index\footnote{The spectral index $\alpha$ is
related to the flux density $S_\nu$ and frequency $\nu$ by
$S_\nu\propto\nu^\alpha$.} $\alpha\simeq -0.3$ to 0.0 and high $(\ga10\%)$
degrees of linear polarization \citep{krh06}. Due to the long synchrotron
cooling time of the radio-emitting particles, a radio PWN can act as a
direct calorimeter to reflect the system's integrated history over a long
period of time. More importantly, radio polarimetry provides a powerful
probe of the magnetic field geometry of a PWN, which is relatively less
studied. High resolution polarization measurements have only been carried
out on a handful of bow-shock systems, e.g.\ IC~443 and the Mouse
\citep{ocw+01,yg05}. In particular, VLA observations of the
Mouse revealed a highly ordered field wrapping around the bow
shock at the apex and extending parallel to the nebular axis downstream
\citep{yg05}, in contrast to the theoretical prediction by \citet{rcl05}.

To generalize our understanding of PWNe, we need to study more examples.
In this paper, we report radio observations of a bow-shock nebula powered
by the energetic pulsar \object{J1509--5850}. This pulsar was
discovered in the Parkes Multibeam Pulsar Survey \citep{kbm+03}. It
has a spin period $P=88.9$\,ms, a high spin-down luminosity $\dot{E}
=5.1\times10^{35}$\ergs\, and a relatively young characteristic
age $\tau_c\equiv P/(2\dot{P}) =1.5 \times10^5$\,yr. As with other
energetic pulsars, it has recently been detected in $\gamma$-rays by the
\emph{Fermi Gamma-ray Space Telescope} \citep{wel+10}. The pulsar 
dispersion measure (DM) of 137.7\,pc\,cm$^{-3}$ suggests a distance
ranging from 2.6 to 3.8\,kpc according to different Galactic
free electron models \citep{tc93,cl02}. Throughout this work, we will
adopt a pulsar distance $d=3\,d_3$\,kpc. Some $\sim4\arcmin$ southwest
of the pulsar, there is an elongated radio source \object{MSC~319.9--0.7}
(hereafter \msc) which was identified by \citet{wg96} as a candidate SNR
based on observations with the Molonglo Observatory Synthesis Telescope
(MOST). The radio source has a size of $7\arcmin \times 2\arcmin$, and
consists of a bright central bulge and a clump in the south.

\citet{hb07} reported an observation of the field of \psr\ with the
\emph{Chandra X-ray Observatory}. The X-ray image revealed a tail-like
PWN extending from the pulsar in the same orientation as \msc, suggesting
that it is a bow-shock nebula. At 3\,kpc, the tail has a physical scale
$\sim5$\,pc, one of the longest X-ray tails ever observed. The PWN
has an X-ray luminosity $L_X \approx 10^{33}$\ergs\ between 0.5 and 8\,keV,
corresponding to a high efficiency $\eta \equiv L_X /\dot{E} \approx 
2\times 10^{-3}$. \citet{hb07} argued that \msc\ is too small to be a SNR
associated with the pulsar, and concluded that it is most likely a
background object. On the other hand, \citet[][hereafter
\citetalias{kmp+08}]{kmp+08} presented a detailed study using the same
\emph{Chandra} data set, and suggested that \msc\ could be the radio
counterpart of the X-ray PWN. They found an average flow speed
$>5000$\kms\ and an equipartition field of a few times $10^{-5}$\,G
in the X-ray--emitting region.

Motivated by these previous studies, we have carried out high resolution
radio imaging using the Australia Telescope Compact Array (ATCA) to
investigate the nature of \msc. Observations and data reduction
are described in \S\ref{s2}. We present the results in \S\ref{s3} and
discuss the physical implications in \S\ref{s4}. Finally, we summarize
our findings in \S\ref{s5}.  

\section{OBSERVATIONS AND DATA REDUCTION}
\label{s2}
ATCA imaging of \msc\ was carried out in two 12\,hr observations
with the EW~367 and 750C array configurations on 2007 Nov 16 and
Dec 1, respectively. To obtain a uniform $u$-$v$ sampling and to
boost the signal-to noise ratio, the longest baselines were
excluded in our analysis, giving array spacings from 46\,m to 750\,m.
The observations were made at 3 and 6\,cm (center frequencies at 8.64
and 4.8\,GHz, respectively) simultaneously with all Stokes parameters
recorded. At each wavelength, the total bandwidth of 128\,MHz was
divided into $32\times4$\,MHz overlapping spectral channels. After
discarding the edge channels and channels affected by self-interference,
and merging the overlapping adjacent channels, a usable bandwidth of
104\,MHz is left, which is split into $13\times8$\,MHz channels. The
entire field was covered by a 9-pointing mosaic patten with pointings
spaced by 3\arcmin, smaller than the FWHM of the primary beam even at 3\,cm.

The flux density scale is set by observations of the primary calibrator,
PKS B1934--638. In addition, a secondary calibrator, PKS~B1511--55, was
observed every 30\,minutes to determine the antenna gains. All our data
reduction was performed with standard techniques using the \emph{MIRIAD}
package \citep{stw95}. We first examined the data carefully to reject outlier
visibility data points and scans during poor atmospheric phase stability. We
then determined gain, bandpass, flux and polarization calibrations for the
observations, and applied these to the source visibilities. Mosaiced images
in each polarization were formed using the multifrequency synthesis
technique \citep{sw94} to improve the $u$-$v$ coverage and with uniform
weighting to minimize the sidelobes. Images in Stokes \emph{I}, \emph{Q}
and \emph{U} were deconvolved simultaneously using a maximum entropy algorithm
\citep[\texttt{PMOSMEM;}][]{sbd99}, and restored with a Gaussian beam of
FWHM $17\arcsec \times 15\arcsec$, corresponding to the diffraction limit
at 6\,cm with the 750\,m baseline. Finally, we convolved the 3\,cm map to
identical spatial resolution as the 6\,cm one. Our final maps in Stokes
\emph{I}, \emph{Q} and \emph{U} at 3\,cm have rms noise of
0.2\,mJy\,beam$^{-1}$, very close to the theoretical level of
0.18\,mJy\,beam$^{-1}$. At 6\,cm, the noise level is 0.15\,mJy\,beam$^{-1}$
in the Stokes \emph{I} map, and 0.1\,mJy\,beam$^{-1}$ in the Stokes \emph{Q}
and \emph{U} images, again consistent with the theoretical value of
0.1\,mJy\,beam$^{-1}$.

Using the Stokes \emph{Q} and \emph{U} images, we generated maps of the
polarized intensity and position angle (PA) with the task \texttt{IMPOL},
which includes corrections for Ricean bias \citep{wk74}. To obtain the
intrinsic PAs of the polarization vectors, we reprocessed the 20\,cm ATCA
data from the Southern Galactic Plane Survey \citep[SGPS;][]{hgm+06} on a
channel-by-channel basis, applied the rotation measure (RM) synthesis
technique \citep{bd05} to map the foreground Faraday rotation, and then
derotated the 3 and 6\,cm polarization vectors accordingly.

In addition to the radio data, we have also processed the 40\,ks
\emph{Chandra} X-ray observation (ObsID
\dataset[ADS/Sa.CXO#obs/03513]{3513}) reported by \citet{hb07}
and \citetalias{kmp+08} for a multi-wavelength comparison. After the
standard pipeline processing, we generated an exposure-corrected image
in the 0.5-7\,keV energy range and applied a smoothing to 6\arcsec\ 
resolution to maximize the signal-to-noise ratio.

\section{RESULTS}
\label{s3}
\subsection{Nebular Morphology}

\begin{figure*}[ht]
\epsscale{1.0}
\plottwo{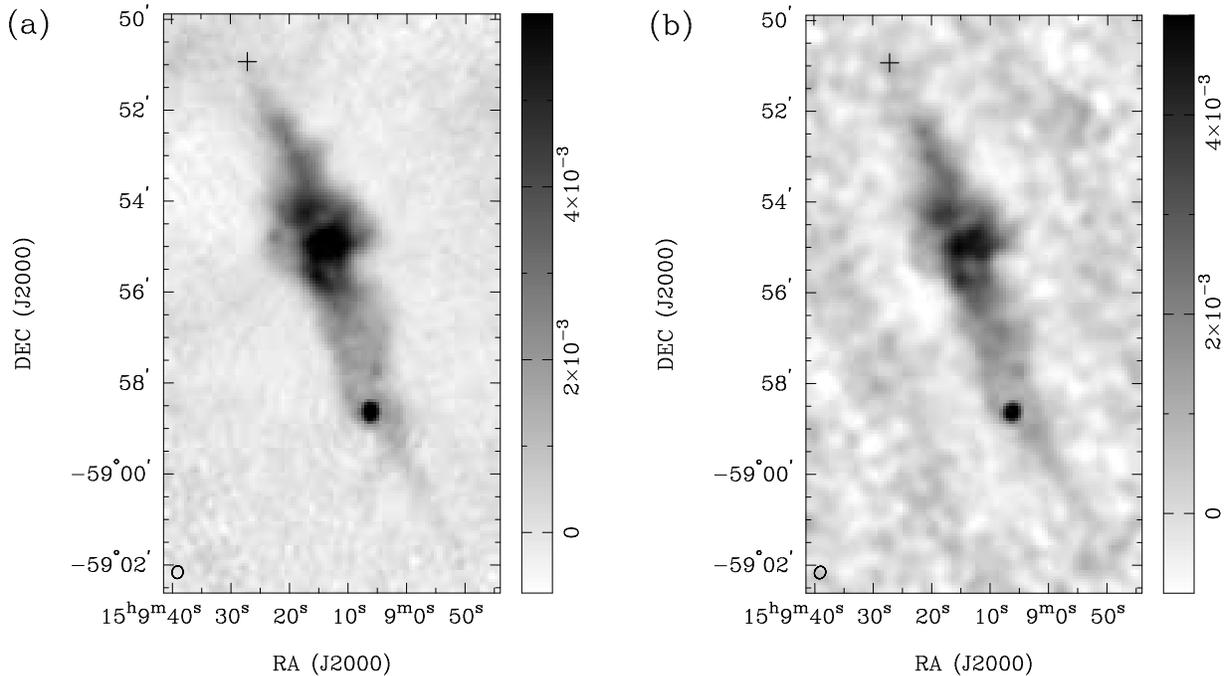}{f1b.eps}
\caption{Radio continuum images of \msc\ at (a) 6\,cm and (b)
3\,cm. The position of \psr\ is marked by the cross. The grey scales
are linear, ranging from --0.7\,mJy\,beam$^{-1}$ to +6\,mJy\,beam$^{-1}$
in the 6\,cm map, and from --0.8\,mJy\,beam$^{-1}$ to
+5\,mJy\,beam$^{-1}$ in the 3\,cm map. The beam size of FWHM $17
\arcsec \times 15\arcsec$ is shown at the lower left of each panel.
\label{radio} }
\end{figure*}

Radio continuum images of \msc\ at 3 and 6\,cm are presented in
Figure~\ref{radio}. The source exhibits a very similar morphology
at both wavelengths; it is highly elongated and extends over 11\arcmin\ 
with a clear symmetry axis along the PA 200\arcdeg\ (measured from north
through east). We note that since the shortest array spacing in the
data is only 46\,m, the sensitivity of the 3\,cm intensity map is
expected to drop for angular scale beyond $\sim3\arcmin$, resulting
in obvious sidelobes in Figure~\ref{radio}(b). \psr\ was not detected in
our observations, but its position, which is marked by the cross in the
figure, coincides with the tip of the nebula. At the pulsar location,
the radio emission is very faint. It gradually brightens towards the
south and shows a cone-like morphology with an opening angle
$\sim35\arcdeg$. The emission peaks at a central bulge 4\farcm5 from
the pulsar and attains a maximum width of 3\arcmin. Beyond that, the
nebula fades and narrows. At 8\arcmin\ southwest of the pulsar, there
is a bright, unresolved radio source with a faint X-ray counterpart
reported by \citetalias{kmp+08}, which could possibly be a
background radio galaxy. Further south, the nebula changes slightly
in orientation ($\sim10\arcdeg$) and narrows to a collimated faint
tail. Finally, beyond $\sim 11\arcmin$ from the pulsar, the nebular
surface brightness drops to near the noise level of the radio maps,
although there is some hint of extension for another 1--2 arcminutes.

\begin{figure*}[!hb]
\epsscale{0.5}
\plotone{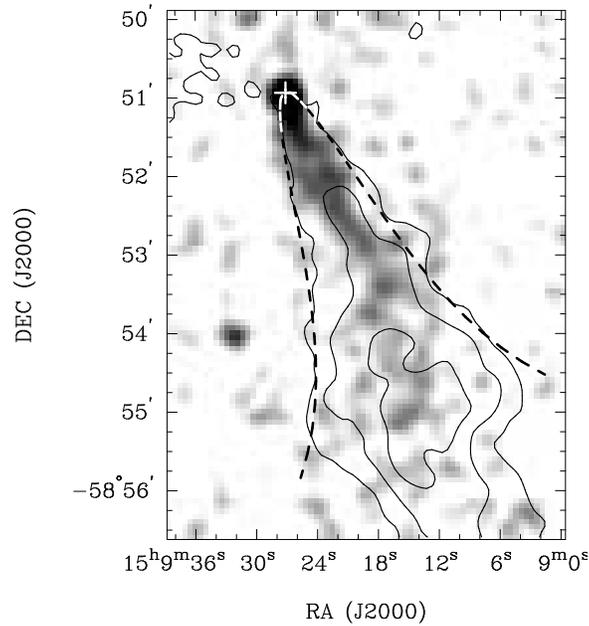}
\caption{Exposure-corrected \emph{Chandra} X-ray image of the field around
\msc\ in the 0.5--7\,keV energy band, smoothed to 6\arcsec\ resolution
and overlaid with the 6\,cm radio intensity contours from Figure~\ref{radio}(a)
at 0.5, 2, and 4.5\,mJy\,beam$^{-1}$. The cross marks the position of
\psr\ and the dotted line shows the best-fit theoretical model given by
\citet{rcl05} (see \S\ref{s43}).\label{cxo}}
\end{figure*}

\label{s31}
Figure~\ref{cxo} shows the exposure-corrected \emph{Chandra} X-ray image,
overlaid with radio contours from the 6\,cm radio intensity map in
Figure~\ref{radio}(a). The X-ray and radio emissions near the pulsar are
well-aligned and are bounded by a common envelope. We followed
\citetalias{kmp+08} by modeling the envelope with a parabola and obtained
$x=3\arcsec \left [ (z+0\farcs 5)/1\arcsec \right ]^{1/2}$, where $x$
and $z$ are the angular distance (in arcseconds) from the pulsar, measured
perpendicular and parallel to the nebular symmetry axis, respectively.
As a comparison, our result is slightly wider than the one reported by
\citetalias{kmp+08}, which may due to the lower resolution of the radio
map than the X-ray image. Figure~\ref{cxo} also indicates an
anti-correlation between the radio and X-ray surface brightness. While the
former increases with distance from the pulsar, the latter peaks at the
pulsar position and fades to the south. At 1\farcm5 downstream from the
pulsar, the X-ray tail suddenly narrows by a factor of 2. We found some
hint of a radio ridge, as indicated by the middle contour in Figure~\ref{cxo},
which may correspond to the narrowed X-ray tail. Further south, the X-ray
emission becomes more diffuse and seems to extend to the radio peak.
Deeper observations are needed to confirm this. Figure~\ref{prof} shows
the surface brightness profiles of the radio and X-ray emissions,
illustrating the features described above.

\subsection{Polarization Properties}
\begin{figure*}[!ht]
\epsscale{0.9}
\plottwo{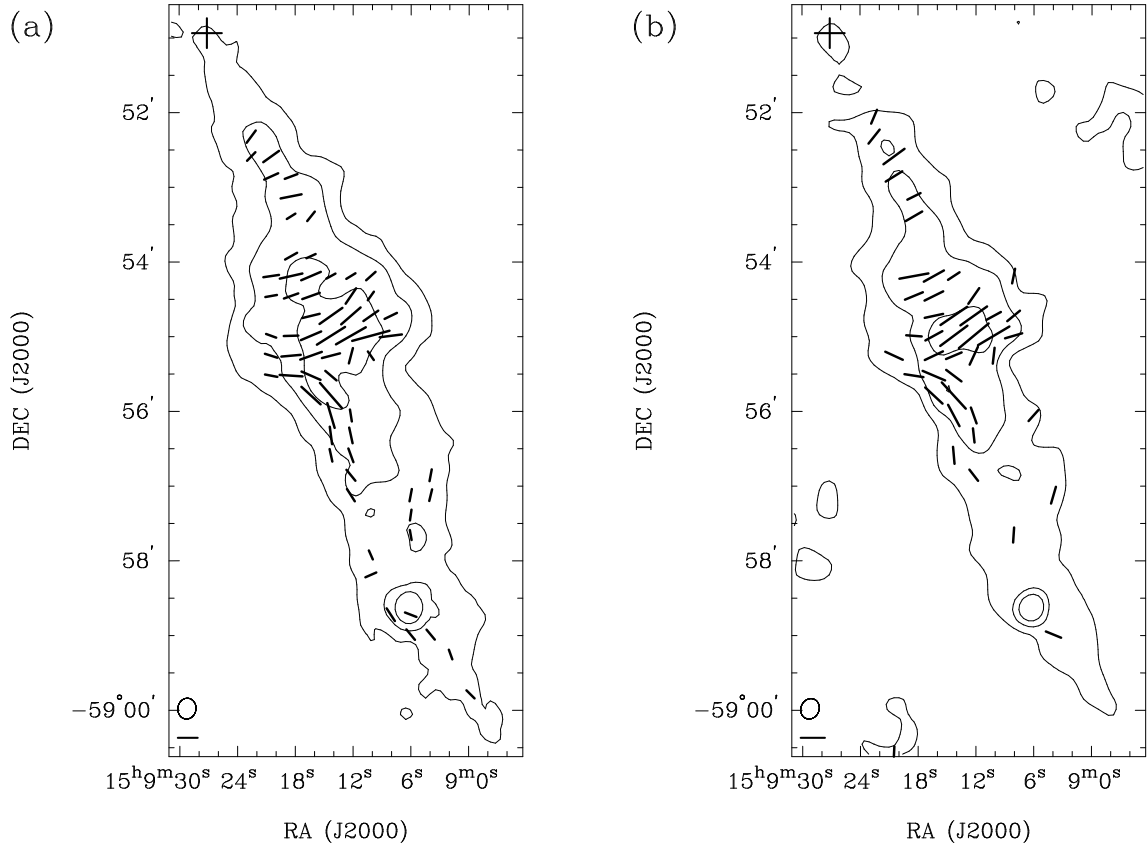}{f4b.eps}
\caption{Intrinsic orientation of polarization $B$-vectors of
\msc\ at (a) 6\,cm and (b) 3\,cm, after correction for the
foreground Faraday rotation, overlaid with the total intensity
contours at the corresponding wavelengths. The 6\,cm map is clipped
at a signal-to-noise ratio (S/N) of 5 for the polarized intensity,
and the 3\,cm map is clipped at a S/N of 3. The contours are at levels
of 0.5, 2, and 4.5\,mJy\,beam$^{-1}$ for the 6\,cm map and 0.6, 2, and
4\,mJy\,beam$^{-1}$ for the 3\,cm map. The vector lengths are
proportional to the polarized intensity, with the scale bar at the
lower left representing 1\,mJy\,beam$^{-1}$. The position of \psr\ is
marked by the cross, and the beam size is shown at the lower left of
each panel.\label{pl}}
\end{figure*}

After correction for the foreground Faraday rotation (see below), the
strength and intrinsic projected orientation of the polarization $B$-vectors
of \msc\ are plotted in Figure~\ref{pl}, revealing a highly ordered magnetic
field structure. The $B$-vectors in the north runs along the northwest-southeast
direction with a PA of 120\arcdeg, nearly perpendicular to the symmetry axis
of the nebula. South of the bulge, the vectors switch direction abruptly,
and show a good alignment with the nebular axis.

The fractional linear polarization map of \msc\ at 6\,cm is shown in
Figure~\ref{frac}. The source is highly linearly polarized and the degree
of polarization appears to be systematically higher around the edges than
in the interior. For instance, it is over 40\% polarized along the southern
edge and $\ga 30\%$ in the north, but only $\la 25\%$ at the central bulge.
The map at 3\,cm, which is not shown here due to low signal-to-noise
ratio, suggests a very similar pattern.

\begin{figure*}[!hb]
\epsscale{0.4}
\plotone{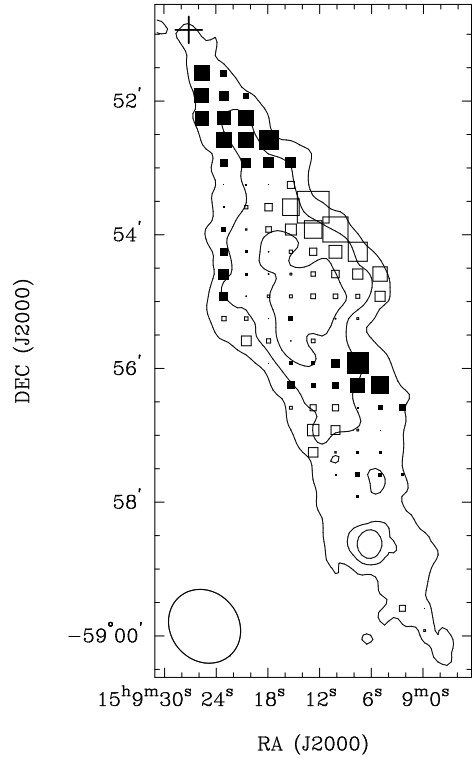}
\caption{Map of rotation measure (RM) towards \msc\ obtained from
the 20\,cm SGPS data, overlaid with the 6\,cm intensity contours
at 0.5, 2, and 4.5\,mJy\,beam$^{-1}$. The filled and open squares
indicate positive and negative RM values, respectively, with the
linear dimensions of each box representing the RM magnitude, ranging
from --100 to +70\,rad\,m$^{-2}$. The cross marks the position of
\psr, and the beam size of the 20\,cm observation (FWHM of $72\arcsec
\times 63\arcsec$) is shown at the lower left. \label{rm} }
\end{figure*}

Figure~\ref{rm} presents the results from RM synthesis of the 20\,cm
data, indicating a small magnitude of RM for most regions. The central
bulge has $|\mathrm{RM}|<20\,$rad\,m$^{-2}$, and the RM slightly increases
to +50\,rad\,m$^{-2}$ towards the tip in the north. The measurement
uncertainties in the map are of a similar order ($\la 50$\,rad\,m$^{-2}$).
Since the compact source in the south is very weakly polarized, we cannot
determine if it has a different RM than the rest of the nebula. Using this
RM map, we derotated the 3 and 6\,cm polarization vectors to their intrinsic
orientation, which are plotted in Figure~\ref{pl}. Due to the small RM
values, the corrections at both wavelengths are generally small, e.g.\ a
RM of 50\,rad\,m$^{-2}$ would correspond to a derotation of 3\arcdeg\ at
3\,cm and 11\arcdeg\ at 6\,cm. As a note, the measurement errors in
the PA of the 3 and 6\,cm vectors are about 8\arcdeg\ and 4\arcdeg,
respectively.

\subsection{Radio Spectrum}
\begin{figure*}[!hb]
\epsscale{1}
\plottwo{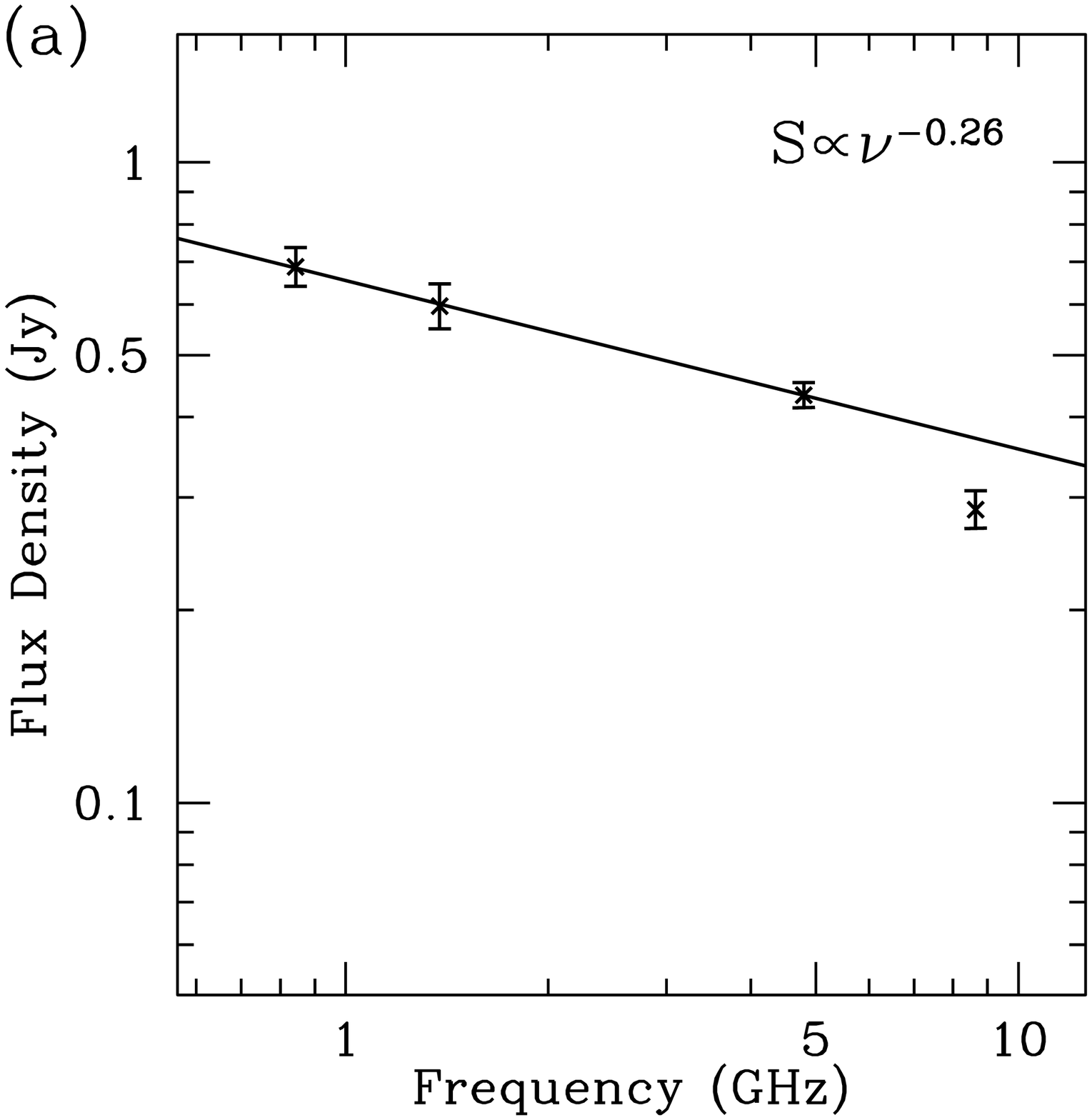}{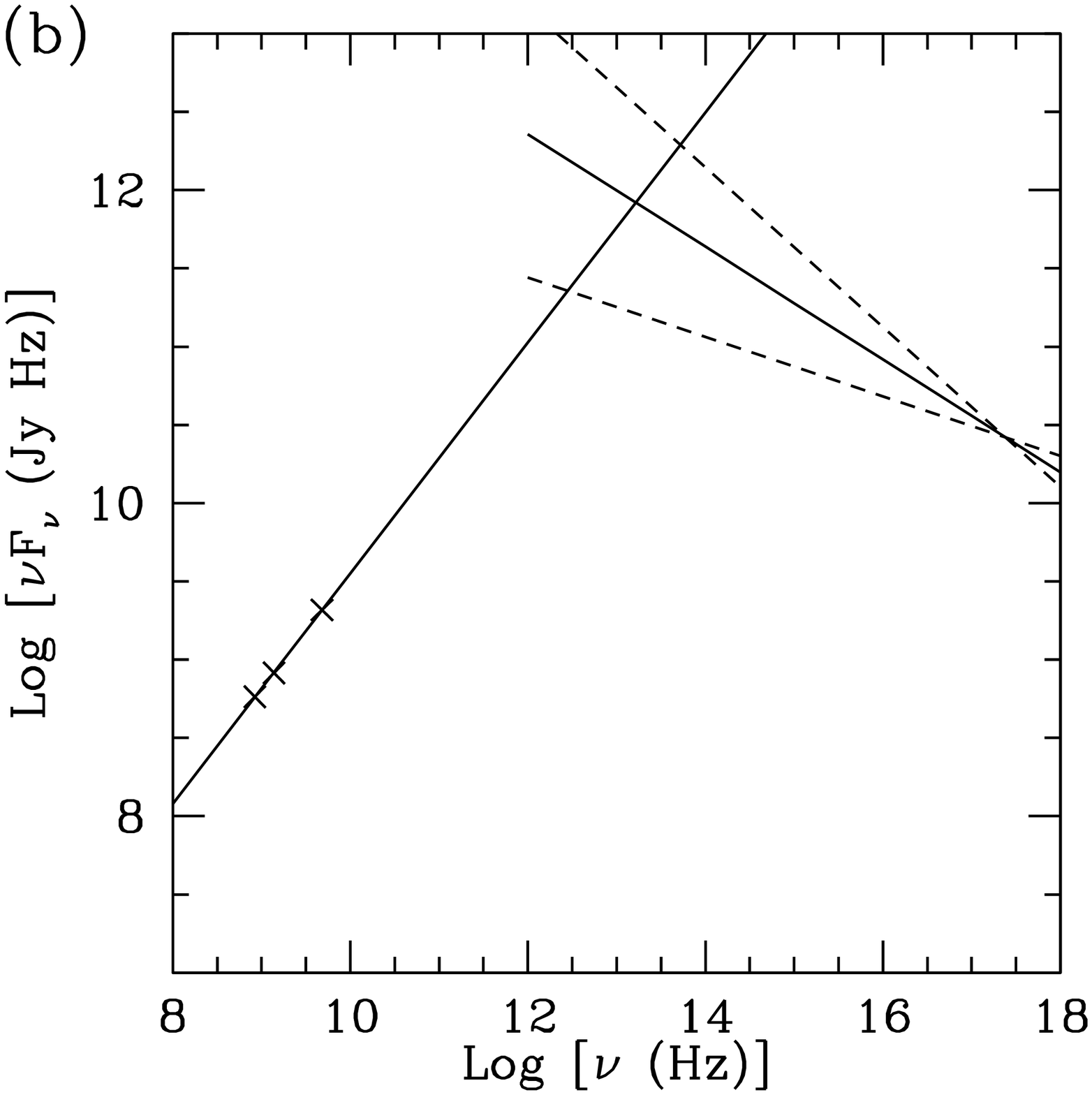}
\caption{(a) Radio continuum spectrum of \msc, after subtracting the
compact source in the south. Due to limited spatial sampling, the flux
at 3\,cm is likely underestimated (see text). Fitting a power-law spectrum
to the other three data points suggests a spectral index $\alpha=-0.26\pm0.04$.
(b) Broadband spectrum of the nebula by comparing to the X-ray results
from \citetalias{kmp+08}. The dotted lines indicate the uncertainty in
the X-ray power-law spectrum.\label{spec} }
\end{figure*}

After subtracting the unresolved source in the south, \msc\ has flux
densities of $0.29\pm0.02$, $0.43\pm0.02$, $0.60\pm0.05$ and
$0.69\pm0.05$\,Jy at 3, 6, 20 and 36\,cm, respectively, as obtained from
our data (3 and 6\,cm), SGPS \citep[20\,cm;][]{hgm+06} and MOST
\citep[36\,cm;][]{wg96}. The resulting radio spectrum is plotted in
Figure~\ref{spec}(a). With the shortest array spacing of 46\,m, the
3\,cm intensity map is only sensitive to angular scales smaller than
2\farcm6. Therefore, the flux measurement at 3\,cm is likely
underestimated and excluded this data point to deduce a spectral index
$\alpha=-0.26 \pm0.04$ from 3 other bands. The compact source has a
substantially steeper radio spectrum of $\alpha=-0.96\pm0.02$, suggesting
that it could be an unrelated background source. Following \citet{gw03},
we resampled the 6\,cm data and compared to the 3\,cm map through
spectral tomography \citep{kr97}. The result indicates no significant
spatial variation in the spectral index along the nebula. Finally, 
a comparison to the X-ray spectrum reported by \citetalias{kmp+08}
is plotted in Figure~\ref{spec}(b). The broadband spectrum suggests a
break near $\sim10^{13}$\,Hz with a change in spectral index $\Delta
\alpha \approx 1$. Integrating the radio spectrum below the break
frequency gives a radio luminosity $L_R=8.7\times 10^{33}$\ergs,
nearly an order higher than the X-ray luminosity.

\section{DISCUSSION}
\label{s4}
\subsection{Nature of \msc}
Our results clearly show that \msc\ is a non-thermal radio source with
a flat spectrum ($\alpha\approx-0.3$) and a high degree of linear polarization
(30--40\%), thereby establishing the PWN nature of the source. Moreover,
the location of \msc\ respect to \psr\ and its alignment with the X-ray
tail strongly suggest that it is a radio counterpart of the bow-shock
nebula. Previous \emph{Chandra} X-ray observation indicates a projected
bow-shock standoff distance $\sim0\farcs5$ \citepalias{kmp+08}, which is
too small to be resolved by our radio maps. Using this value, the
theoretical shape of a bow-shock \citep[e.g.][]{wil96} is about 5 times
narrower than the parabola we obtained. This is likely the result of
deceleration in the flow downstream (see \S\ref{s42} below).

Due to long synchrotron cooling timescale, the radio emission of a PWN
traces the pulsar motion over a large distance, reflecting the path of the
system. Therefore, in the following discussion, we call \msc\ a radio
`trail' rather than a `tail', and its orientation implies a pulsar proper
motion direction towards the northeast at PA $\sim20$\arcdeg.
At the pulsar distance $3\,d_3$\,kpc, the radio trail has a physical
extent over $10\,d_3$\,pc. For a space velocity $\sim300$\kms\ 
\citepalias{kmp+08}, if the pulsar is nearly as old as its spin-down age
$\tau_c=1.5\times 10^5$\,yr, then it would have traveled a distance 8
times longer than the radio trail. Even accounted for the projection, we
think it is unlikely that the radio PWN reflects the entire history
of the system, and the pulsar birth site could be much further south. At
4\arcdeg\ southwest of the pulsar, there is an open cluster Ruprecht~112
\citep[C1453--623;][]{abr+70}
lining up exactly with the nebular axis. However, this cluster is poorly
studied and the distance is unknown. If this is the birth site of the pulsar,
the large angular separation would require a high pulsar velocity of
1400\kms, an older age, or a closer source distance. A direct measurement
of the pulsar proper motion could provide more insights into the birth site.

The small RM towards the source is not unexpected, since the line of
sight to the source is near the interarm region between the Crux and
Norma Galactic spiral arms, where the RM contributed by the ISM is small
\citep{hml+06}. As their work suggests, our result is consistent with a
source distance below 4\,kpc. The ratio between the PWN RM
($\la 20$\,rad\,m$^{-2}$) and the pulsar DM
\citep[137.7\,pc\,cm$^{-3}$;][]{kbm+03} indicates a small mean field along
the line of sight
\begin{equation}
<B_\parallel>=1.232\,\frac{\rm RM}{\rm DM}\,\mu\mathrm{G}
\la0.2\,\mu\mathrm{G} \ .
\end{equation}

Finally, we note that the compact source in the south has a flux density
of 0.1\,Jy at 20\,cm. \citet{hac+03} found that radio sources with at
least this flux level have a number density of 2.0\,per square degree in
the sky. Therefore, within the PWN area $\sim0.01$\,deg$^2$, the
chance probability of having such a background source is about 2\%,
not completely negligible.

\subsection{Magnetic Field Strength and Flow Structure of \msc}
Figure~\ref{spec}(b) shows that the X-ray synchrotron radiation of the
PWN could extend down to $10^{13}$\,Hz without a spectral break. According
to \citetalias{kmp+08}, this value implies an equipartition field strength
$\sim100\,\mu$G near the pulsar and a projected flow speed
$v_{\mathrm{flow,X}} \sim 4\times 10^5$\,km\,s$^{-1}$. To estimate
the magnetic field downstream beyond the X-ray tail, we consider the
radio emission in the southern half of the nebula beyond 4\farcm5 from
the pulsar. We integrate the spectrum in the range $10^7-10^{13}$\,Hz
to obtain a luminosity $L=4.4\times 10^{33}d_3^2$\ergs\ 
(after excluding the background source), with the corresponding
emission volume $V=2.0\times 10^{57} d_3^3\,$cm$^3$. This gives
an equipartition field
\begin{equation}
B_\mathrm{eq}=\left [6\pi c_{12}(1+k)L\Phi^{-1}V^{-1} \right ]
^\frac{2}{7} \,\mathrm{G} \approx20\,d_3^{-\frac{2}{7}}
(1+k)^\frac{2}{7} \Phi^{-\frac{2}{7}} \, \mu\mathrm{G} \ , 
\end{equation}
where $k$ is the ion to electron energy ratio, $0<\Phi<1$ is the volume
filling factor, and $c_{12}$ is a constant weakly depends on the spectral
index \citep{pac70}.\footnote{Although there is recent revision to this
formula, e.g.\ \citet{bk05}, we note that the revised version does not work
for flat spectrum objects in which $\alpha > -0.5$.} Assuming $k=0$ and
$\Phi=1$, we obtained $B_\mathrm{eq}= 20\,\mu$G, a few times lower than
that of X-ray--emitting region. We note that this result is independent
of the lower frequency limit, and insensitive to the upper limit,
e.g.\ the range $10^7-10^{11}$\,Hz yields $B_\mathrm{eq}=15\,\mu$G.

\label{s42}
While the flow speed in the radio PWN cannot be determined precisely
without knowing the maximum energy of the synchrotron radiation, the
absence of X-ray emission can still provide some rough estimates. From
the point where the X-ray tail vanishes, the radio nebula extends
7\arcmin\ further downstream, corresponding to a physical length
$l\sim6d_3$\,pc. Assuming most photons in the region have
frequencies $\nu<10^{16}$\,Hz ($\approx0.04$\,keV, nearly the
detection threshold of \emph{Chandra}), the synchrotron cooling time scale 
\begin{equation}
\tau_\mathrm{syn}> 2000\left( \frac{B}{20\mu\mathrm{G}} \right)^{-
\frac{3}{2}} \left(\frac{\nu} {10^{16}\mathrm{Hz}}\right)
^{-\frac{1}{2}}\, \mathrm{yr} \end{equation}
suggests a flow speed 
\begin{equation}
v_\mathrm{flow,R} \sim \frac{l}{\tau_\mathrm{syn}} < 3000\,d_3
\left(\frac{B}{20\mu\mathrm{G}}\right)^{\frac{3}{2}} \left(\frac{\nu}
{10^{16}\mathrm{Hz}}\right)^{\frac{1}{2}} \mathrm{km\,s^{-1}} \ . 
\end{equation}
Here we ignored adiabatic cooling, since
the nebula cross-section shows no widening beyond the central bulge.
Moreover, if the majority of particles radiate below the break
frequency of $10^{13}$\,Hz, the flow speed will then be as low as
$\sim100$\,km\,s$^{-1}$. We emphasize that these estimates are
highly uncertain because the exact spectrum is unknown. Nonetheless,
a comparison to the X-ray results suggests a significant deceleration
of the flow when moving downstream, with the speed dropping by 1--2
orders of magnitude.

\subsection{Magnetic Field Geometry and Physical Interpretation}
\label{s43}
The most remarkable feature of \msc\ is the distribution of the
polarization vectors. Since synchrotron emission of a PWN is optically
thin and Faraday thin (i.e.\ depth depolarization is negligible) at our
observed wavelengths, the polarization maps in Figure~\ref{pl} represent
the magnetic field of the nebula projected on the sky plane. This
implies a helical field structure in the north, with the axis of
symmetry parallel to the nebular axis, and hence to the pulsar's
inferred direction of motion. This is the first time an azimuthal
field geometry has been seen in a bow-shock nebula. Axisymmetric toroidal
fields about pulsar rotation axes have been observed in young PWNe
within SNRs \citep[e.g.][]{dlm+03,hes08}, and have been successfully
reproduced by MHD simulations \citep[see,][and references therein]{vda+08}.  
As an analogy to these systems, the field configuration of \msc\ could
be understood if the pulsar motion aligns with its spin axis. Such an
alignment is not uncommon \citep{jhv+05}, and could be the result of a
momentum kick at the birth of a neutron star \citep[see,][]{sp98,nr07}.
If this is the case for \psr, the jet-like polar outflow would contribute
to the X-ray tail. The MHD instabilities (e.g.\ kink and sausage
instabilities) in the flow could then help explain the nonuniform surface
brightness in the tail downstream \citep{kp08}. Future polarization
measurements of the pulsar profile may reveal the projected orientation of
its spin axis, directly confirming the above picture. However, since
\psr\ has a relatively low degree of polarization \citep{wj08}, this may
require the next generation of radio telescopes.

Further support for the spin-velocity alignment is given by analytic
studies of pulsar wind bow-shocks. \citet{rcl05} showed that an aligned
pulsar can wind up its magnetic field to form an azimuthal field in the
magnetotail behind, possibly resembling what we found in \msc. Assuming
energy equipartition between the field and the particles, these authors
derived an analytic expression for the magnetotail radius $r_m(z)$ in
terms of the distance $z$ from the star \citep[Eq.\ 34--35 in][]{rcl05}.
Using the bow-shock standoff distance of 0\farcs5 \citepalias{kmp+08},
their model predicts a too narrow nebular morphology. If the model
parameters are allowed to vary, we are able to obtain a better fit,
which is overplotted in Figure~\ref{cxo}, although it requires a very
large standoff distance $z_{\rm sh}\approx r_m(0)=2.0 \times10^{17}
d_3$\,cm corresponding to 4\arcsec, even at the lower limit
5000\,km\,s$^{-1}$ of the flow speed \citepalias{kmp+08}. The fit also
suggests a large deceleration in the flow of length scale $L=3$\,pc,
much shorter than the total distance traveled by the pulsar, as proposed
in their model. This could explain why the observed nebular morphology
is much wider than the theoretical predictions.

\begin{figure*}[!ht]
\epsscale{0.4}
\plotone{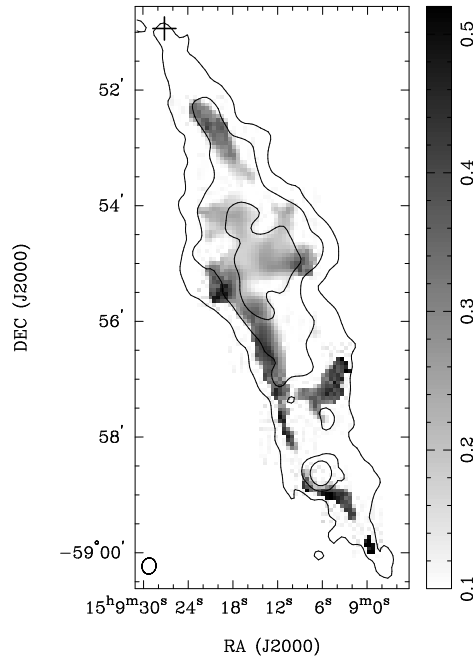}
\caption{Fractional linear polarization map of \msc\ at 6\,cm,
clipped at a signal-to-noise ratio of 5 for the polarized
intensity, overlaid with the total intensity contours at levels
of 0.5, 2, and 4.5\,mJy\,beam$^{-1}$. The grey scales are linear,
ranging from 10 to 52\%. The position of \psr\ is marked by the
cross and the beam size is shown at the lower left.
\label{frac}}
\end{figure*}

Beyond the central bulge, the magnetic field of \msc\ switches
direction abruptly and runs parallel to the nebular axis, similar to
that of the Mouse PWN reported by \citet{yg05}. The change in field
geometry seems difficult to understand. One possible scenario could
be a two-component field structure consisting of a poloidal core
surrounded by a toroidal field, such as the one suggested by
\citet{klr+08} for DA~495\footnote{We should note that the field
structure of DA~495 is inferred from the RM distribution of the PWN.
However, internal Faraday rotation is generally not expected for
non-thermal plasmas such as pulsar winds.}.
If the toroidal field decays faster downstream, this would leave the
poloidal core component, resulting in a switch in the overall field
orientation. In this picture, the projection of the two field components
leads to a lower degree of polarization in the PWN interior than
around the edge as observed. Near the transition region where the two
field strengths are comparable, we expect the total polarization
to cancel out. However, such a complete depolarizing region is not
observed in Figure~\ref{frac}. Indeed, Figure~\ref{pl} shows that the
polarization vectors near the bulge have intermediate orientation,
suggesting physical change of the field direction rather than
projection effect. Therefore, we conclude that this scenario is less likely.

Hinted by the flow deceleration and the relatively lower degree of
polarization at the bulge, an alternative physical scenario is that the
plasma flow might become unstable downstream, then drives an expanding bubble
into the ISM, similar to the model suggested by \citet{vi08} to explain
the H$\alpha$ observations of the Guitar Nebula, the bow-shock PWN powered
by PSR~B2224+65 \citep{crl93}.
If we decompose the magnetic field into random and ordered components,
then the relative strength $q^2=\case{<b^2>}{<B^2_\perp>}$ between the
random field $b$ and the ordered field perpendicular to the
line of sight $B_\perp$ can be estimated from the degree of
polarization $p$. For the case of energy equipartition, 
\begin{equation}
\frac{p}{p_0}= \frac{1+\case{7}{3}q^2}{1+3q^2+ \case{10}{9}q^4}\ , 
\end{equation}
where $p_0=\frac{3-3\alpha}{5-3\alpha}$ is the intrinsic degree of
polarization \citep{bk05}. As indicated in Figure~\ref{frac}, the nebular
emission is $\sim40\%$ polarized around the edge, implying a highly
ordered field of $q^2\approx 1.2$. On the other hand, the lower degree
of polarization $p\sim25\%$ in the interior gives $q^2\approx 3.2$.
This indicates a substantial random field, suggesting some hints of
turbulence, which could be due to flow instability. If this is the case,
then the actual value of $q^2$ would be even higher due to the intrinsic
anisotropy of MHD turbulence \citep{bss+03}.

Assuming continuous energy injection and adiabatic expansion,
\citet{vi08} give the angular radius $\theta_{\rm b}$ of the bulge in
terms of the instability angular scale $\lambda$ using the Sedov-Taylor
solution:
\begin{equation}
\theta_{\rm b}=f_{\rm b}\left (\frac{4\pi c}{v_\ast} \right)^{1/5}
\theta_0^{2/5} \left( \theta_{\rm d}-\lambda \right )^{3/5} \ ,
\end{equation}
where $f_{\rm b}$ is a function of order unity, $v_\ast$ is the pulsar
velocity, $\theta_{\rm d}$ is the angular separation between the bubble
center and the pulsar, and $\theta_0$ is the angular size of the
bow-shock standoff distance \citep[see Eq.~4 in][]{vi08}. Our radio
maps in Figure~\ref{radio} indicate $\theta_{\rm b} =80\arcsec$ and
$\theta_{\rm d}=280\arcsec$, together with $\theta_0=0\farcs5$ and
$v_\ast\approx 400$\,km\,s$^{-1}$ from \citetalias{kmp+08}; these
suggest $\lambda\approx170\arcsec$, implying a relatively slow growth
rate of the instability. The ratio $\lambda /\theta_0\approx 340$
is higher than the value 80 inferred from the Guitar Nebula, but
still within a few times of the instability growth length scale
$\lambda_{\rm g}\lesssim 60\,\theta_0$ as discussed by \citet{vi08}.
A more realistic estimate with the magnetic field taken into account
could lead to a lower expansion rate, and hence a slightly smaller
$\lambda$. The estimate above predicts an expansion rate 
\begin{equation}
\mu_{\rm b}\approx \frac{2}{5}\frac{v_\ast}{d} \frac{\theta_{\rm b}}
{\theta_{\rm d} -\lambda} = 8\,\mathrm{mas\,yr^{-1}} 
\end{equation}
for the bulge, too small to be detected by the ATCA over a reasonable
time span, and the emission is too faint for VLBI measurements.

For the orientation switch of the polarization vectors downstream, we
speculate that the expanding bubble may channel particles
into the low-pressure cavity evacuated by the pulsar wind along the path,
thus changing the field configuration. Finally, we should point out that
this picture fails to explain how the polarization structure is preserved
during the bubble expansion. Further simulation work is thus necessary to
complete the study.

\subsection{Multiwavelength Comparison}
\begin{figure*}[!hb]
\epsscale{0.5}
\plotone{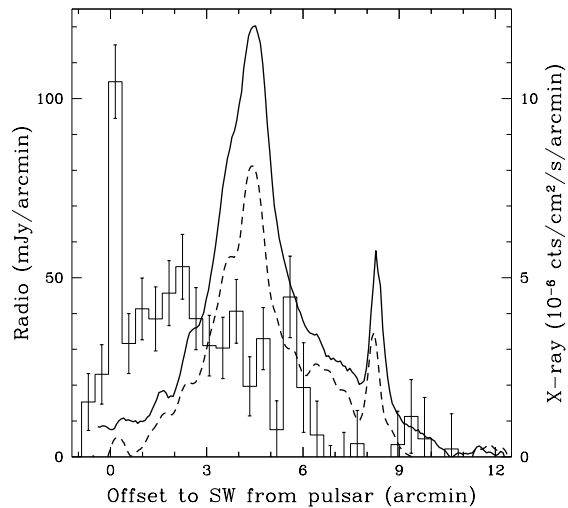}
\caption{Linear brightness profiles of \msc\ in radio and X-rays, extracted
from 3\arcmin\ wide rectangular regions along the nebular axis. The dotted
and solid lines represent the radio profiles obtained from the 3 and
6\,cm intensity maps in Figure~\ref{radio}, and the histogram is extracted
from the exposure-corrected \emph{Chandra} X-rays image binned to
25\arcsec.\label{prof}}
\end{figure*}

Figure~\ref{prof} show that the X-ray tail is brightest near the pulsar
while the radio emission is extremely faint in the same region.
\citetalias{kmp+08} extracted the X-ray counts from a
$14\arcsec\times24\arcsec$ ellipse at the head of the PWN (excluding the
pulsar), and obtained an X-ray power-law spectrum with index $\alpha=
-0.8\pm0.3$. If the absorption column density is allowed to vary,
they found $\alpha=-0.27\pm0.25$, identical to the radio spectral index
of the overall PWN. Using the same extraction region, we obtained flux
densities of $0.5\pm0.3$\,mJy and $0.6\pm0.2$\,mJy at 3 and 6\,cm, respectively,
from the radio intensity maps in Figure~\ref{radio}. Given the X-ray spectrum,
the faintness in radio emission is not unexpected. If we extrapolate the
two different X-ray power-laws to radio wavelengths without any spectral
breaks, we expect flux densities of 0.5\,$\mu$Jy--14\,mJy at 3\,cm and
0.6\,$\mu$Jy--20\,mJy at 6\,cm, consistent with the observed values, albeit
not very constraining. To determine if there is any intrinsic break in the
injection spectrum, further studies will require deeper radio and X-ray
observations, as well as flux measurements at wavelengths in between
(e.g.\ infrared).

Moving downstream from the pulsar, while the X-ray tail fades due to
synchrotron burn-off, the radio emissivity gradually increases. As
\citetalias{kmp+08} point out, this could be caused by the deceleration
of the flow, and similar spatial radio--X-ray anti-correlations have been
observed in other PWNe with mildly relativistic jet-like polar outflows,
e.g.\ the ones powered by PSRs B1509--58, B1706--44 and Vela
\citep{gak+02,rnd+05,kp04}. For the case of \msc, the bulge is $\sim15$
times brighter than the tip from the 6\,cm linear profile (Figure~\ref{prof})
and 10 times wider (Figure~\ref{radio}). Therefore, the volume emissivity
at the bulge is 15\% of that at the tip. As discussed above, the flow speed
drops by 10--100 times between the two regions. After accounted for
the geometry, this corresponds to a factor of 1--10 decrease in particle
number density $n$ downstream. Due to the strong dependence of synchrotron
emissivity on the field strength ($L/V \propto n\, B^{7/2}$ for energy
equipartition, or $ \propto n\, B^2$ if the particle distribution remains
unchanged), our results indicate a minimal field decay between the tip and
the bulge, at most by a factor of 2. While this seems incompatible with the
field strengths estimated in \S\ref{s42}, it argues that energy
equipartition is not achieved near the pulsar and the actual magnetic
field could be close to $20\,\mu$G at the tip of the PWN.

\subsection{Comparison with Other Radio Bow-shock PWNe}
For a handful of bow-shocks with detailed radio studies, their radio
emissions generally peak at the pulsar position and show good spatial
correlations with the X-ray counterparts, e.g.\ the Duck, the Mouse and
IC~443 \citep{kgg+01,gvc+04,gcs+06}. The clear radio--X-ray anti-correlation
of \msc\ likely reflects a different flow condition for the source compared
to these other systems. Additionally, an azimuthal $B$-field geometry has
never previously been found in bow-shocks. In the following discussion we
will focus on the Mouse PWN, since it has high resolution polarimetric
observations that allows a direct comparison with our study.

The magnetic field of the Mouse wraps around the bow-shock at the apex,
it then switches orientation abruptly behind the pulsar and runs parallel
to the nebular axis downstream \citep{yg05}. Since the radio emission is
very faint at the tip of \msc, it is unclear if a similar field direction
switch may exist near the pulsar. The azimuthal field geometry in \msc\ 
is not found in the Mouse. This could reflect different spin-velocity
alignments of the central pulsars in the two systems. While we suggested
that \msc\ could be an aligned case, we may expect a completely different
field geometry if the misalignment angle is large. Future MHD simulations
with anisotropic winds and pulsar orientations taken into consideration
could help verify this picture.

More intriguing is the field structure at large distances from both pulsars.
Figure~\ref{pl} shows that the field orientation of \msc\ exhibits a
90\arcdeg\ switch near the central bulge, at $\sim4.5\,d_3$\,pc from
the pulsar. Similar behavior is also found in the Mouse: the polarization
vectors change direction abruptly at several places along the nebular
axis, with the most prominent one at $\sim5\,d_5$\,pc downstream from
the pulsar, and the behavior appears to be quasi-periodic \citep{yb87}.
This could be due to MHD instabilities in the flows, but the detailed
mechanism is not well understood. More examples are necessary to determine
whether this is a common feature among bow-shocks with long trails.

\section{CONCLUSIONS}
\label{s5}
We have presented a detailed radio polarization study of \msc\ at
3 and 6\,cm. Our results confirm that this source is a bow-shock PWN
associated with \psr, and reveal its central bulge morphology,
which could represent an adiabatically expanding bubble drove
by the flow instability. We found an equipartition field strength
of $\sim 20\,\mu$G and a flow speed $<3000$\kms\ in the radio nebula,
indicating a substantial deceleration of the flow as compared to the
X-ray--emitting region near the pulsar. The polarization measurements
reveal a remarkable azimuthal field geometry in the northern half of
the PWN, which qualitatively agrees with the theoretical prediction for
an axisymmetric bow-shock. As compared to the well-studied case of the
Mouse PWN, \msc\ shows very different field structure and surface
brightness distribution, illustrating the diversity in the physical
properties of bow shocks.

To conclude, our study demonstrates that high resolution radio
polarimetry can provide a powerful diagnostic tool for probing the
physical conditions of a PWN. Future works on expanding the sample
of polarimetric observations on bow-shock nebulae are essential
to understand the general properties of the population.

\acknowledgements
We thank the referee for useful suggestions. We are grateful to Chris
Hales for help with the ATCA observations, and to Oleg Kargaltsev and
K.S.\ Cheng for discussions. The Australia Telescope is funded
by the Commonwealth of Australia for operation as a National Facility
managed by CSIRO. C.-Y.N.\ and B.M.G.\ acknowledge the support of the
Australian Research Council through grant FF0561298.

{\it Facilities:} \facility{ATCA ()}


\medskip

\clearpage

\end{document}